\documentclass[final,1p,times]{elsarticle} % do not change
\usepackage{graphicx} % do not change
\usepackage{amssymb} % do not change
\usepackage{amsthm} % do not change
\usepackage{lineno} % do not change

\journal{Nuclear Physics A} % do not change
\begin{document} % do not change

\begin{frontmatter} % do not change

%% QM09Author: please enter your  
%% Title, author and address info here; please do not use footnotes

% Your Title - please insert
\title{Measurements of high-$p_T$ probes in heavy ion collisions at CMS}

% Principle author, and co-authors - please insert
\author{G. I. Veres $^{a}$ on behalf of the CMS collaboration}
\address[a]{CERN, Geneva, Switzerland}

%%-------------------------------------------------------
\begin{abstract} % do not change
The capabilities of the CMS detector at the LHC will be described for 
measuring high-$p_T$ hadrons, photons and jets in heavy ion collisions. 
Detailed simulations of various studies planned with the CMS 
apparatus, including charged particle tracking,
jet reconstruction using calorimetry, dimuon and isolated photon 
detection and the measurement of in-medium fragmentation functions using 
high-$p_T$ photon-jet correlations will be discussed. 
\end{abstract} % do not change

\end{frontmatter} % do not change
%% QM09: we keep linenumbers at least for initial version
%\linenumbers % do not change

%%-------------------------------------------------------------------

%\section{Introduction}\label{intro}

High-$p_T$ processes have proven to be an important tool in investigating the hot, 
dense matter created in the collision of relativistic heavy ions. 
The large suppression observed for high-$p_T$ hadronic yields 
in central heavy ion collisions relative to the binary scaling of p+p collisions is 
evidence of large partonic energy loss in the medium.
%At RHIC, the medium-induced energy loss is so large that it is difficult to discern 
%the transport properties of the medium either from high-$p_T$ spectra or from 
%back-to-back hadron correlations. Because of the increased jet cross-section at the 
%LHC, measurements of photon-jet or $Z^0$-jet correlations become possible. 
At RHIC energies, several complications have arisen in the interpretation of the 
high-$p_T$ phenomena. The high-$p_T$ hadrons (and jets) are surface-biased if the 
created medium is opaque, masking the real amount of medium-induced parton energy 
loss \cite{surfacebias}. Fragility of some observables (like nuclear modification 
factors) hinders their discriminative power between models and model 
parameters \cite{fragility}.

In the new energy frontier at the LHC a larger $p_T$ region, fully developed 
jets, and several new observables will be accessible. New ways that will be
available to study energy loss 
include $\gamma$-jet, $Z^0$-jet, dijet correlations, and jet fragmentation function 
measurements. Since the photon (or~$Z^0$) escapes from the medium with little 
interaction, it gives a calibrated probe of the original parton energy.
The first heavy ion run at the LHC is scheduled for late 2010 at 
$\sqrt{s_{_{\rm NN}}}\approx$ 4~TeV. 

%\section{Capabilities of CMS to measure hard probes}\label{cms}

The CMS apparatus \cite{cms_paper} is equipped with various subsystems that make it a powerful 
tool to study high-$p_T$ phenomena in heavy ion collisions. The highly segmented 
electromagnetic and hadronic calorimeters have a large longitudinal coverage up to 
$|\eta|<6.6$, including the CASTOR calorimeter \cite{castor}, and the
Zero Degree Calorimeter will detect spectator neutrons at $|\eta|>8.3$.
The muon detectors will be used to reconstruct $Z^0$, $J/\psi$, $\Upsilon$ 
particles, with coverage of $|\eta|<2.4$. The silicon tracker system can reconstruct 
charged tracks with good efficiency and purity. The silicon pixel layers have less 
than 2\% hit occupancy even in heavy ion collisions. The high level trigger will be 
able to inspect each event before the trigger decision, including 
complicated on-line reconstruction algorithms \cite{certalk}.
Some more details on these capabilities will be given below.

\begin{figure}[t]
\centering
\includegraphics[width=0.415\textwidth]{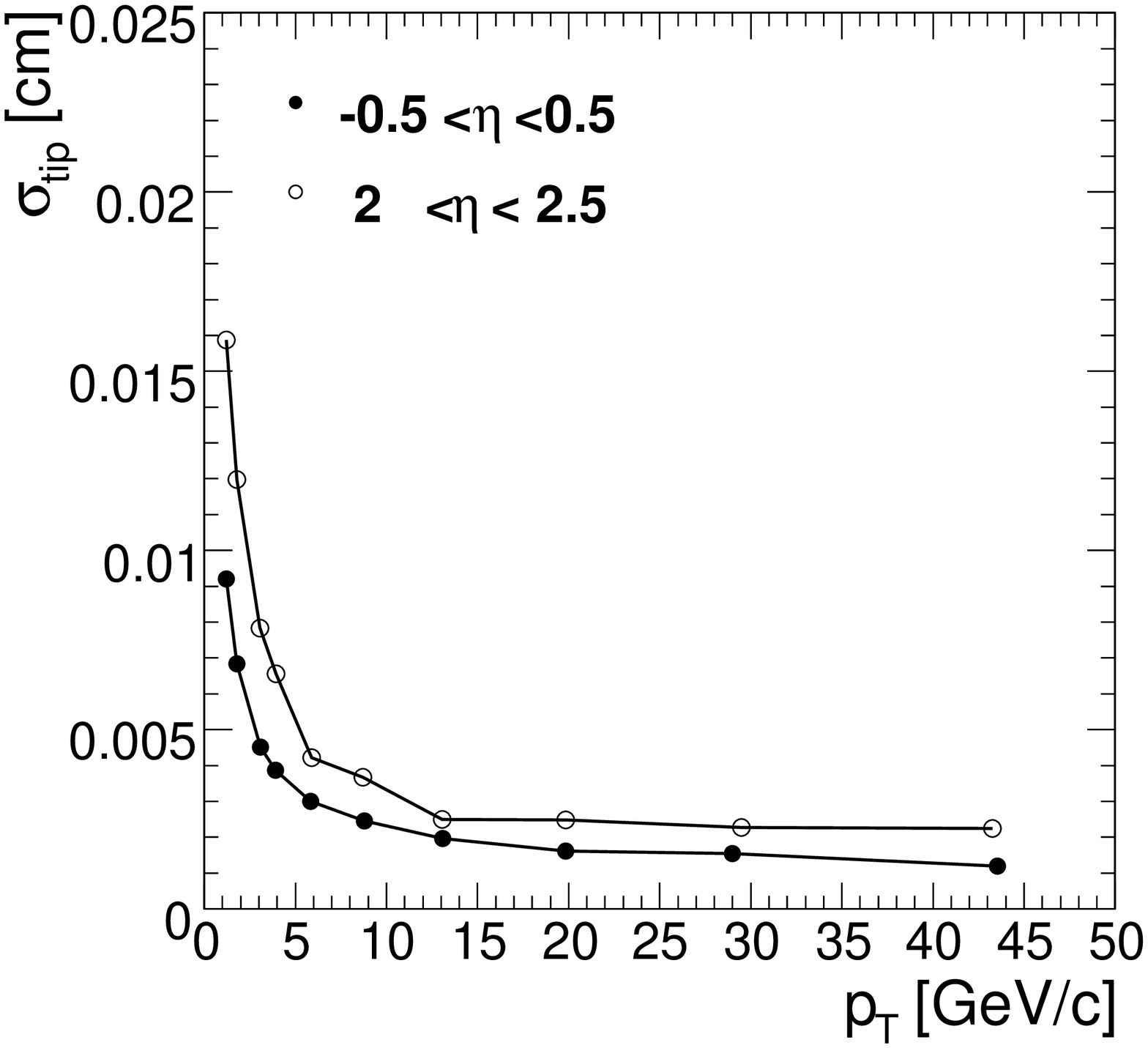}
\includegraphics[width=0.445\textwidth]{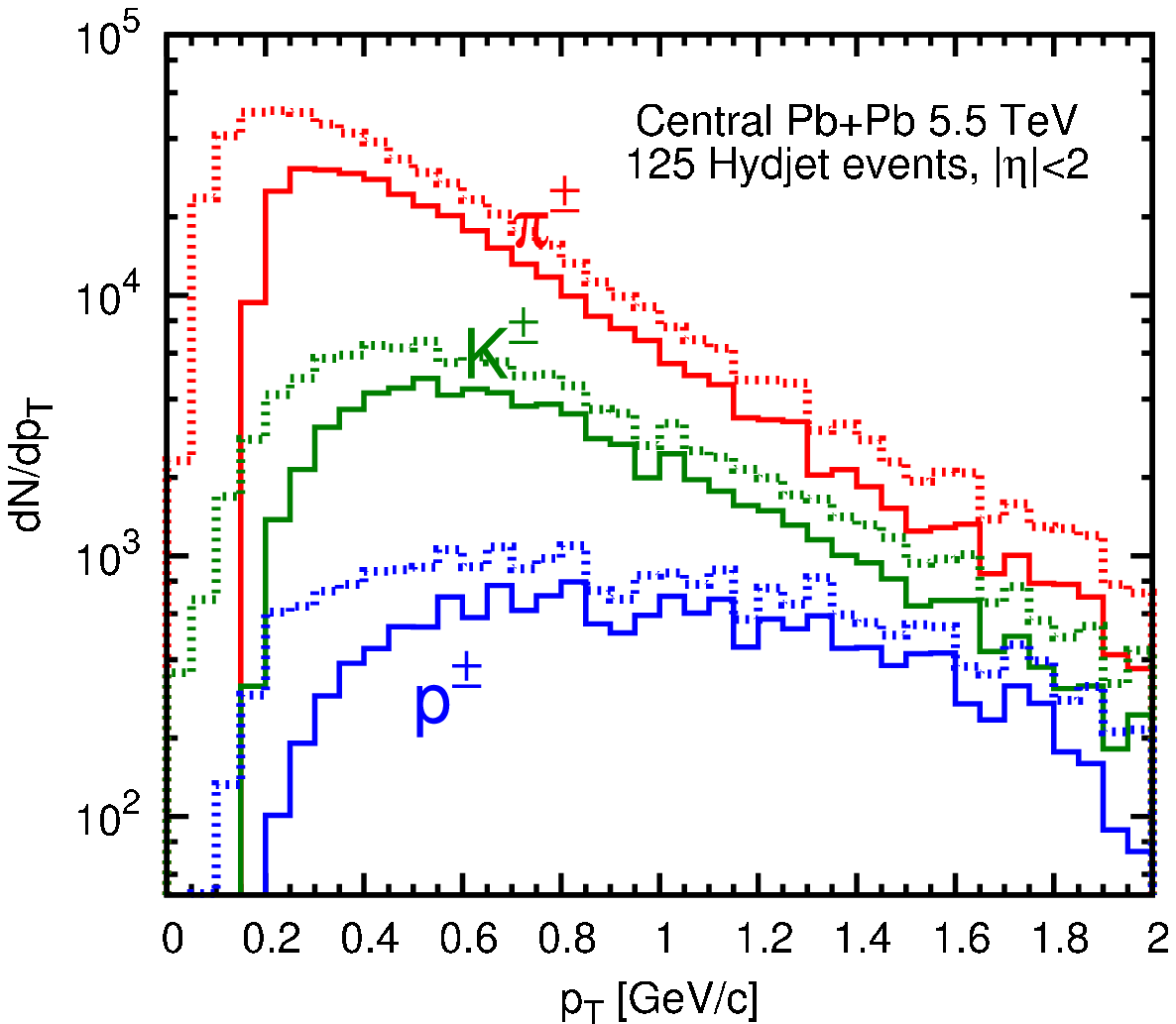}
\caption[]{
Left: $p_T$ dependence of the transverse impact parameter resolution achieved in 
heavy ion events (with $dN_{ch}/dy|_{y=0} = 3200$) in the barrel (full symbols) and 
in the forward endcap (open symbols) regions.
Right: reconstructed (solid lines) and generated (dotted lines)
$p_T$ distributions of the pions, kaons and protons produced in 125
central Pb+Pb HYDJET events at 5.5~TeV/nucleon.}
\label{tracking}
\end{figure}

%% \subsection{Tracking} %----------------------------------------------

The tracking performance of the CMS detector was studied under a 
conservative assumption on the charged particle multiplicity for central Pb+Pb 
events, at $dN_{ch}/dy|_{y=0}=3500$~\cite{cer_tracking}. The simulation results show  
75\% reconstruction efficiency and less than 5\% fake rate in the $1<p_T<30$ GeV/c 
range for charged particles. The $p_T$ resolution was found to be 1-3\% depending on 
$\eta$ and $p_T$.
The left panel of Fig.~\ref{tracking} shows the expected transverse impact 
parameter resolution of charged tracks that will become relevant for displaced 
vertex reconstruction~\cite{cer_tracking}. The $p_T$ reach of the tracking in CMS was 
extended down to $p_T\approx$ 200 MeV/c, and particle identification capability based 
on the specific energy loss (dE/dx) in the silicon tracker was demonstrated 
\cite{hitdr}. The reconstructed and simulated $p_T$ spectra of 
identified hadrons are also shown in Fig.~\ref{tracking} (right).

%% \subsection{Dimuons} %----------------------------------------------

\begin{figure}[h]
\centering
\includegraphics[height=0.38\textwidth]{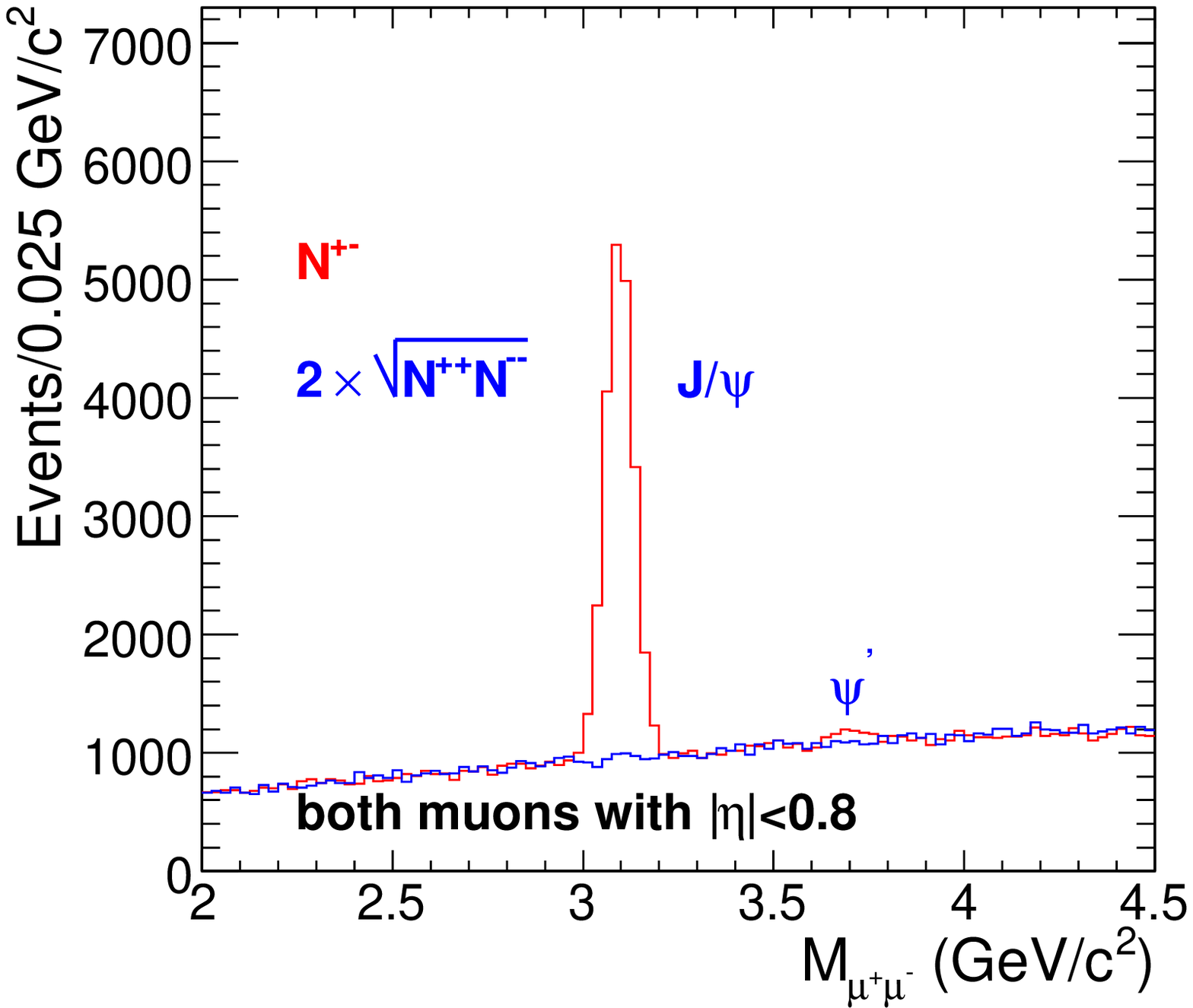}
\includegraphics[height=0.38\textwidth]{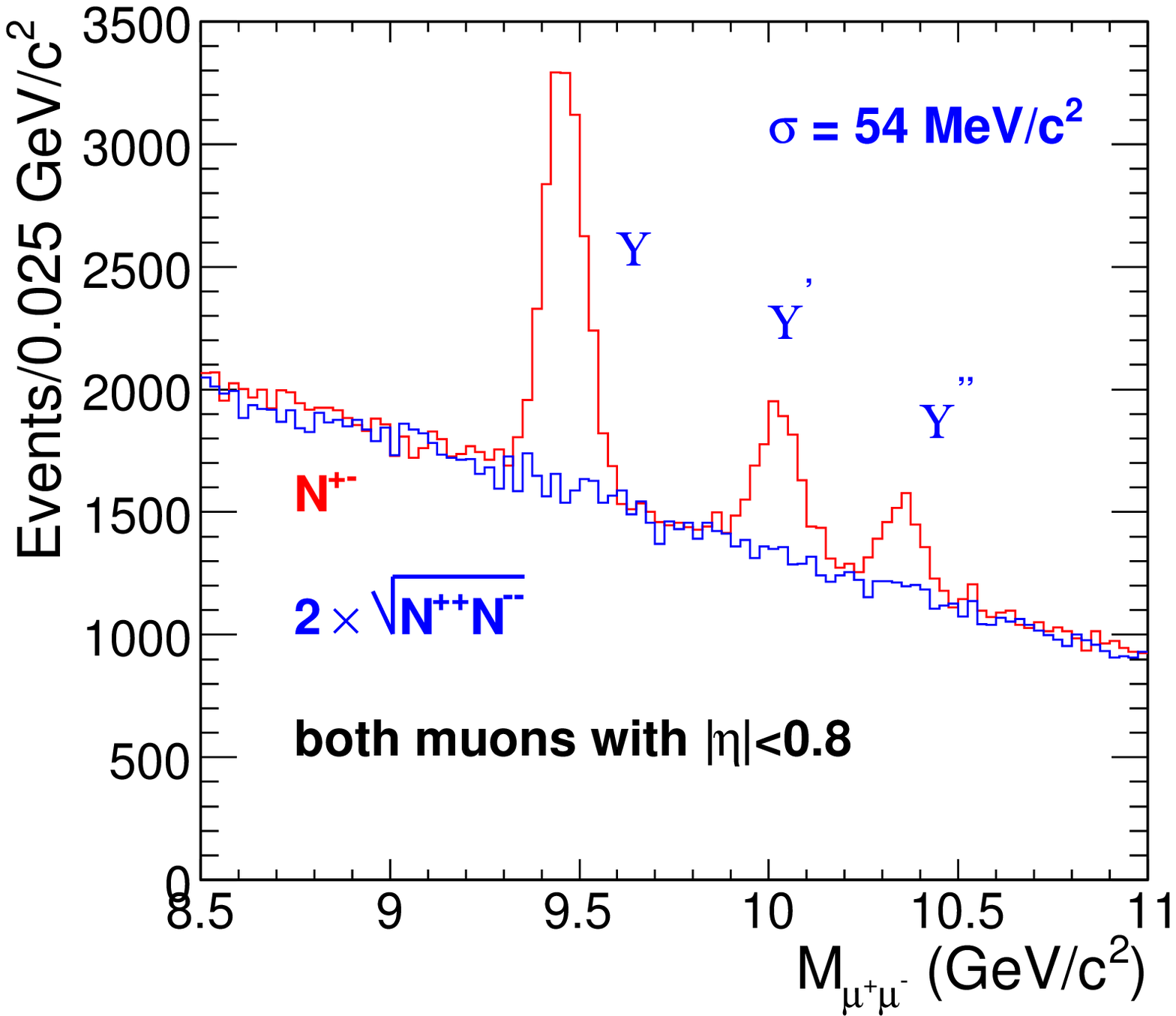}
\caption[]{
Reconstructed invariant mass spectra of opposite-sign and like-sign muon pairs from
simulated central Pb+Pb events with $dN_{ch}/d\eta|_{\eta=0}$ = 2500, in the $J/\Psi$
(left) and $\Upsilon$ (right) mass regions, where both muons have $|\eta|<0.8$.}
\label{qqbar}
\end{figure}

The CMS detector features excellent dimuon reconstruction capabilities,
due to the high, 4~Tesla magnetic field and accurate tracking and muon detectors.
Figure \ref{qqbar} shows the reconstructed dimuon mass spectra in the $J/\Psi$  
(left) and $\Upsilon$ (right) dimuon mass regions for simulated central Pb+Pb 
events with $dN_{ch}/d\eta|_{\eta=0}$ = 2500. The statistics correspond to
$0.5 nb^{-1}$ integrated luminosity.
The signal to background ratio is about 5 for the $J/\Psi$ and 1 for the
$\Upsilon$ where both muons have $|\eta|<0.8$. A mass resolution of
54 MeV/$c^2$ can be achieved for the $\Upsilon$ states~\cite{hitdr}.

%% \subsection{Jets} %----------------------------------------------

High energy jets can be reconstructed in the CMS calorimeters using the iterative 
cone algorithm with event-by-event background subtraction. Although the 
determination of the 
absolute jet energy scale is difficult, the jet finding efficiency is 
more than 50\% at $E_T=50$~GeV and close to 100\% above 70 GeV~\cite{irina_jets}.

The $E_T$ resolution is about 16\% at $E_T=100$~GeV with a background of 
$dN_{ch}/dy|_{y=0}$ = 5000. 
The typical angular resolution of the jet axis obtained 
is $\sigma_\phi=0.03$ and $\sigma_\eta=0.02$.
By applying a series of jet triggers in a nominal $0.5 nb^{-1}$ run, the statistical 
$p_T$ reach of the jet and charged hadron spectrum measurement is 500 and 300 GeV/c, 
respectively~\cite{certalk,jet_trig}. The high statistics of reconstructible jets 
opens the way to more detailed jet quenching studies.

\begin{figure}[t]
\centering
\includegraphics[width=0.58\textwidth]{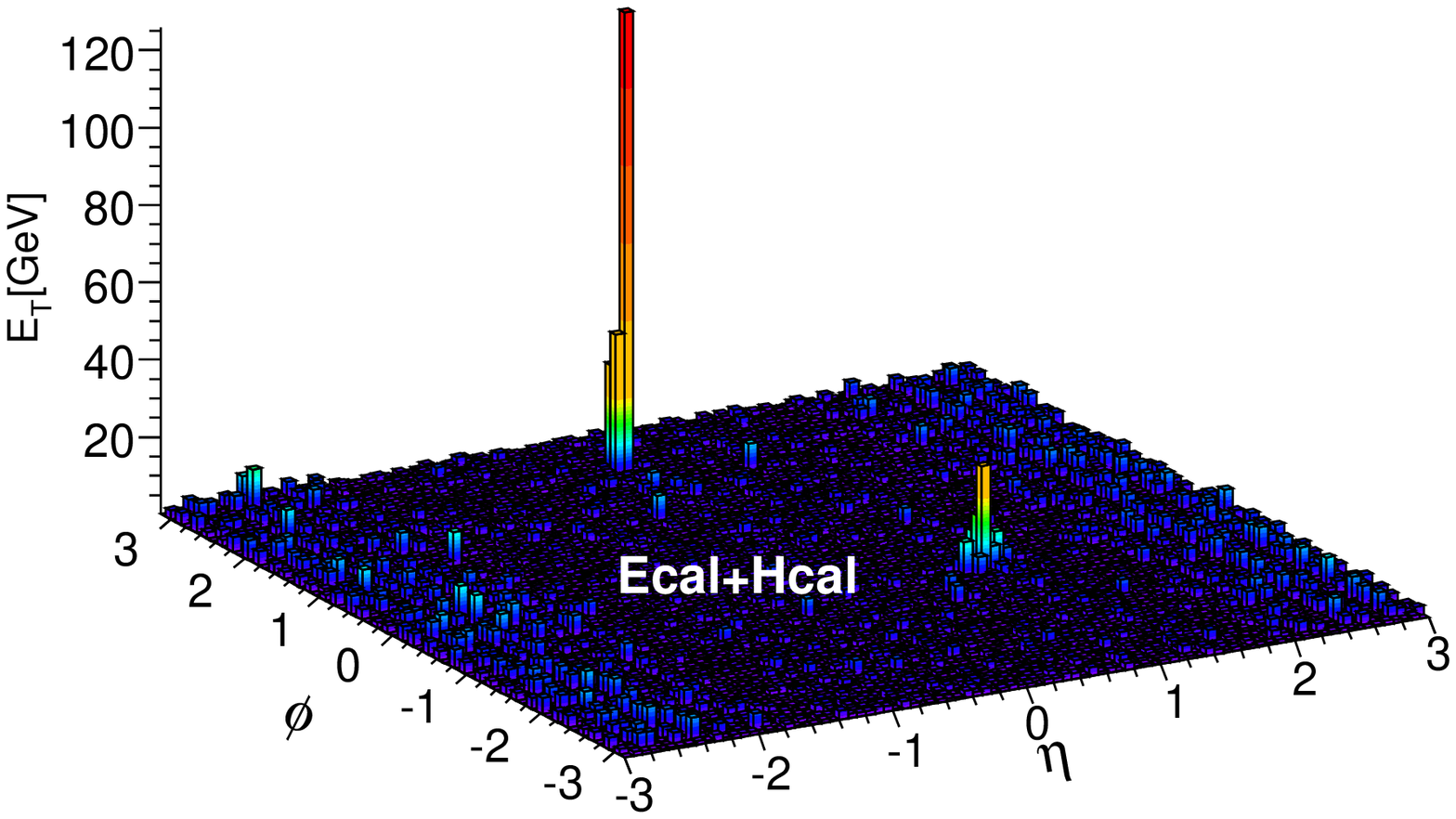}
\includegraphics[width=0.4\textwidth]{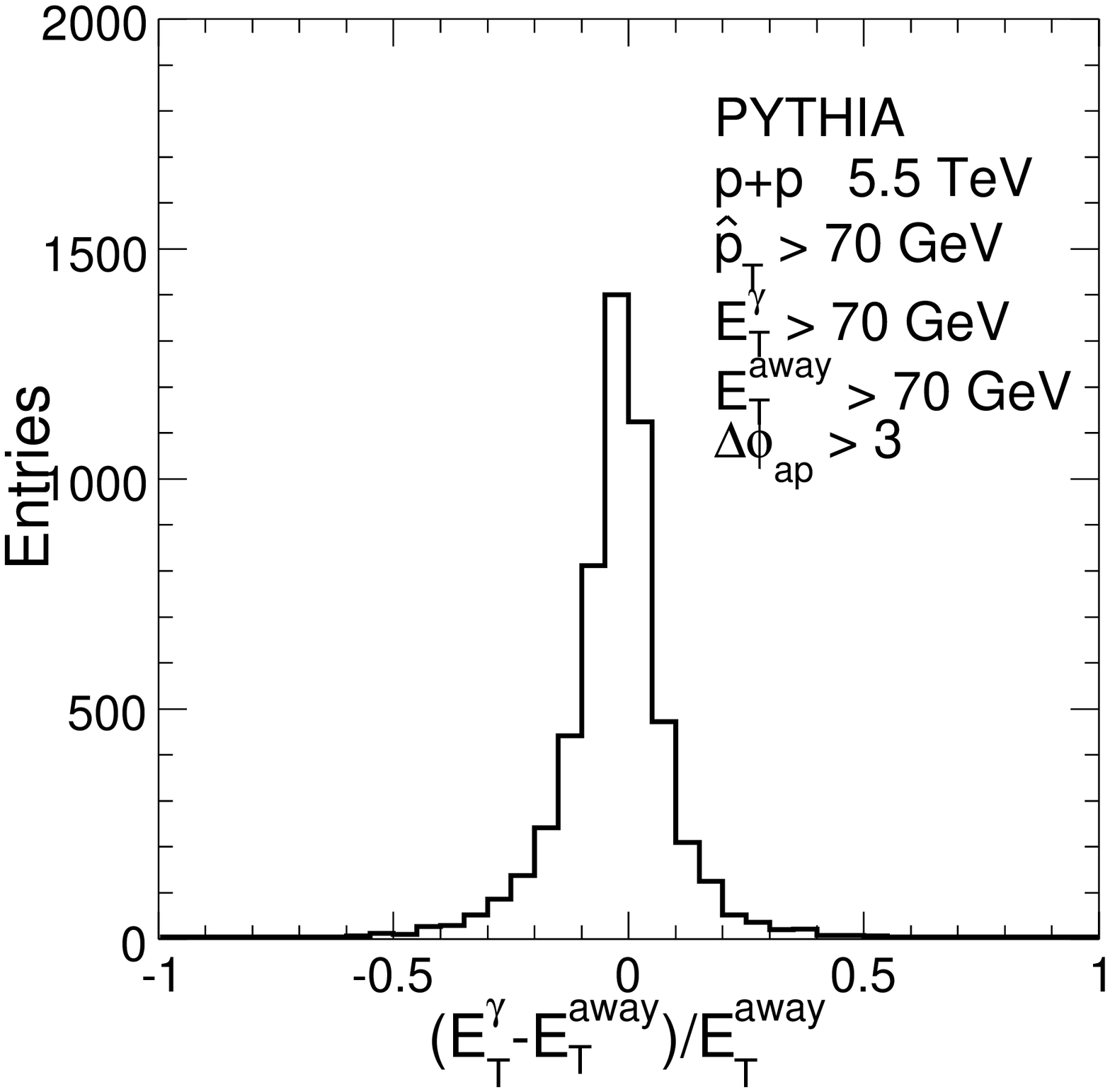}
\caption[]{
Left: energy response on the $\eta-\phi$ plane for a simulated p+p event 
producing a $\gamma$-jet final state embedded in a Pb+Pb event at 
$\sqrt{s_{_{\rm NN}}}=5.5$ TeV. The quenched jet is the cluster at negative $\phi$.
Right: the balance between the gamma and the away side parton $E_T$ using the 
PYTHIA event generator.}
\label{gammajetbalance}
\end{figure}

%% \subsection{Gamma-Jet} %----------------------------------------------

The high resolution electromagnetic calorimeter with large coverage and segmentation 
makes it possible to analyze $\gamma$-jet events, and use the measured $\gamma$ 
energy to calibrate the jet energy scale, coincidentally measuring the properties 
of the quenched away side jet. Figure \ref{gammajetbalance} shows a 
simulated $\gamma$-jet event 
as seen by the calorimeters, and the relatively tight correlation between the $E_T$ 
of the $\gamma$ and that of the away side parton at the event generator level 
\cite{gammajet_pas}.
Isolated photons are selected for this analysis using an isolation cut based on a 
combination of various cluster shape variables, suppressing $\pi^0$-s produced in 
jets usually associated with large hadronic activity. 

\begin{figure}[h]
\centering
\includegraphics[width=0.410\textwidth]{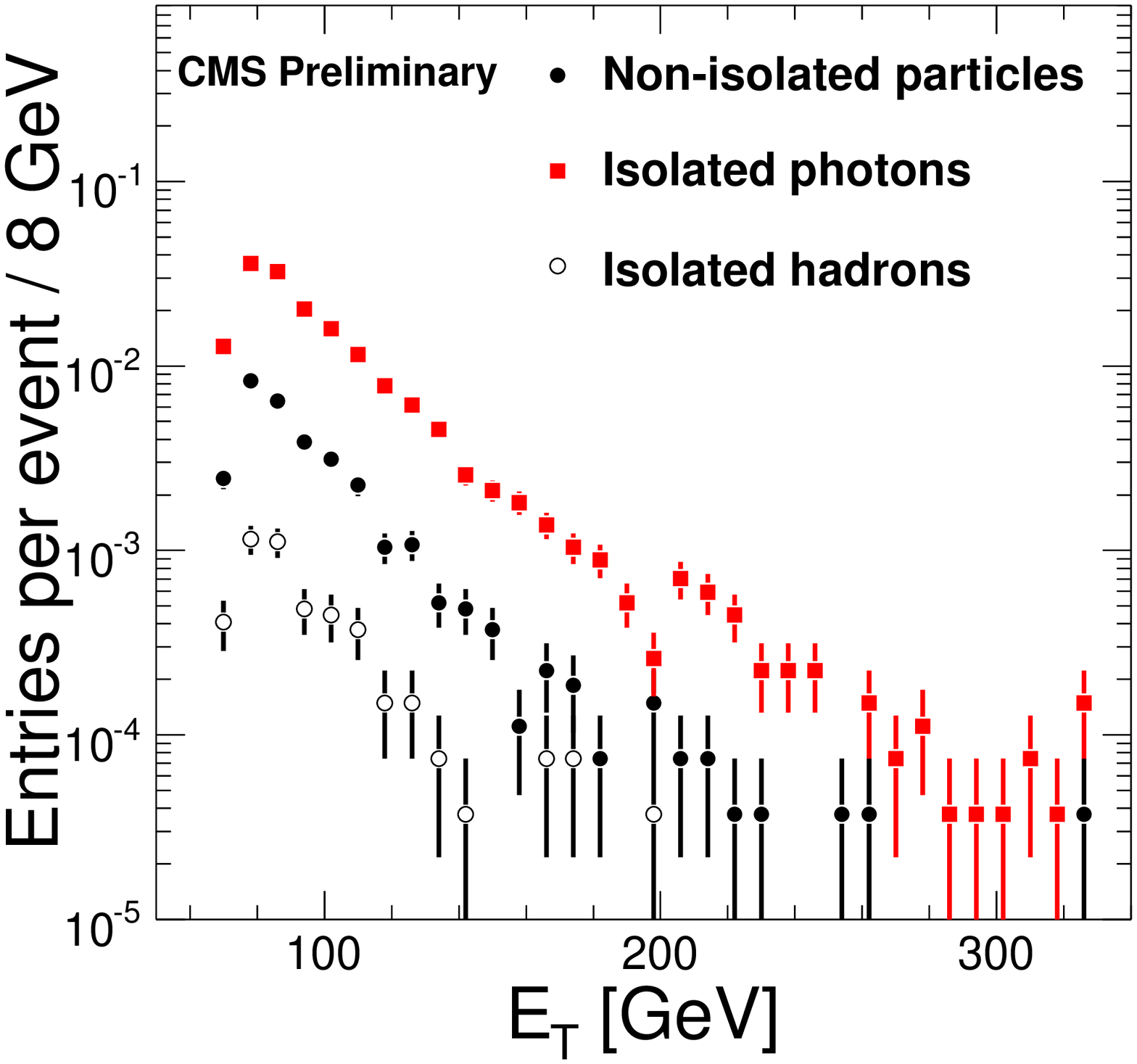}
\includegraphics[width=0.407\textwidth]{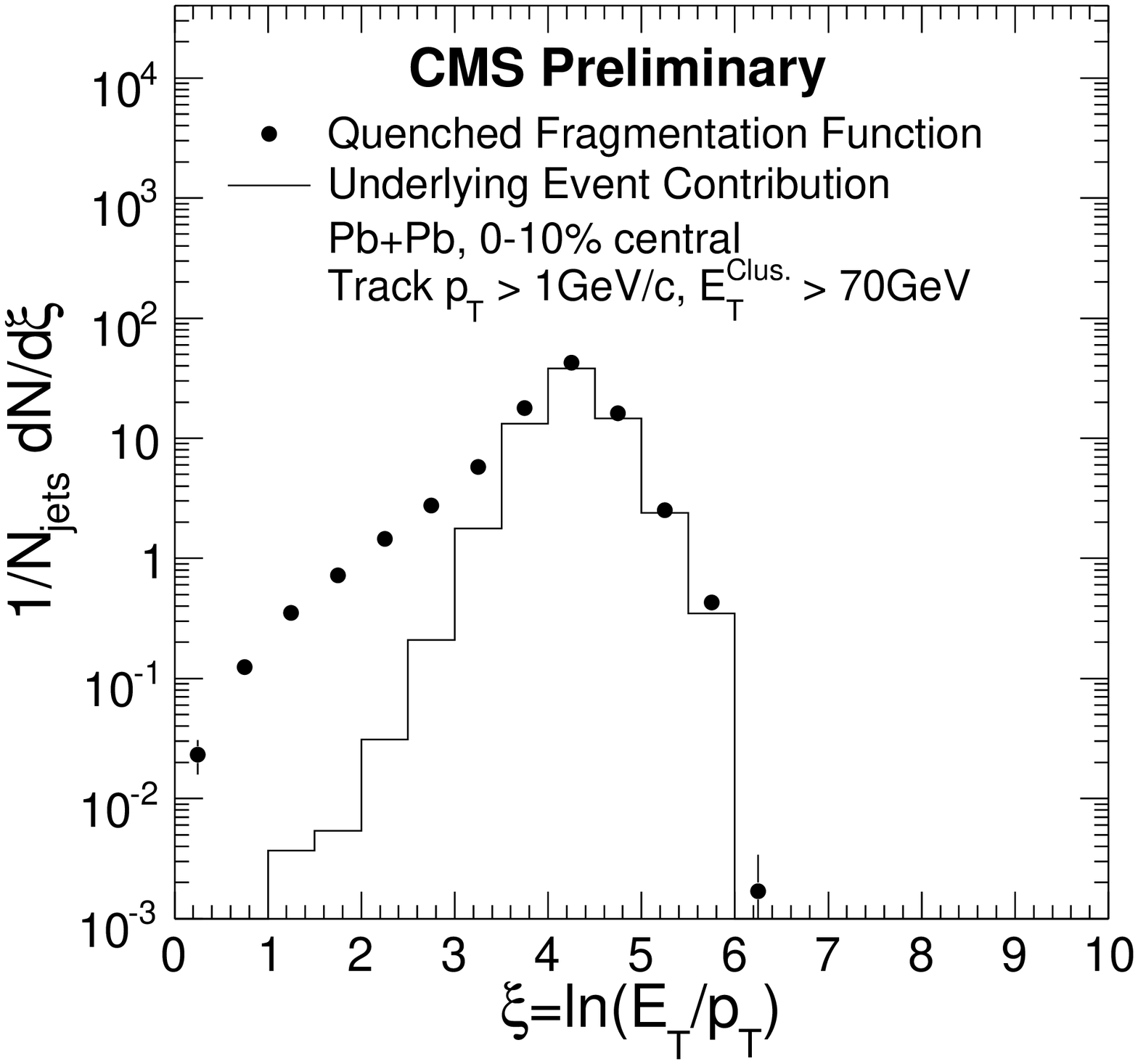}
\caption[]{Left: the $E_T$ distribution for isolated photons, hadrons, and 
non-isolated particles that pass the photon isolation cut, according to simulation.
Right: fragmentation function of quenched high energy jets (symbols) including the 
contribution of the underlying event (histogram) at low $p_T$.}
\label{photoniso}
\end{figure}

A signal to background ratio 
of 4.5 for isolated photons is achieved, as shown on the left 
panel of Fig.~\ref{photoniso}. As a next step, the away-side jet was 
reconstructed, on the opposite side of the $\gamma$, requiring $E_T>30$~GeV.
The fragmentation function, $dN/d\xi$ where $\xi=ln(E_T/p_T)$, was obtained from 
the charged particle tracks that were reconstructed within the jet cone 
of radius R=0.5. Finally, the 
contribution from the underlying event with soft hadrons, shown on the right panel 
of Fig.~\ref{photoniso}, was subtracted. The expected statistics for a 0.5~$nb^{-1}$ 
heavy ion run at $\sqrt{s_{_{\rm NN}}}=5.5$~TeV is 4300 events with $E_T^\gamma>70$~GeV and 1200 events with 
$E_T^\gamma>100$~GeV. 

\begin{figure}[t]
\centering
\includegraphics[height=0.374\textwidth]{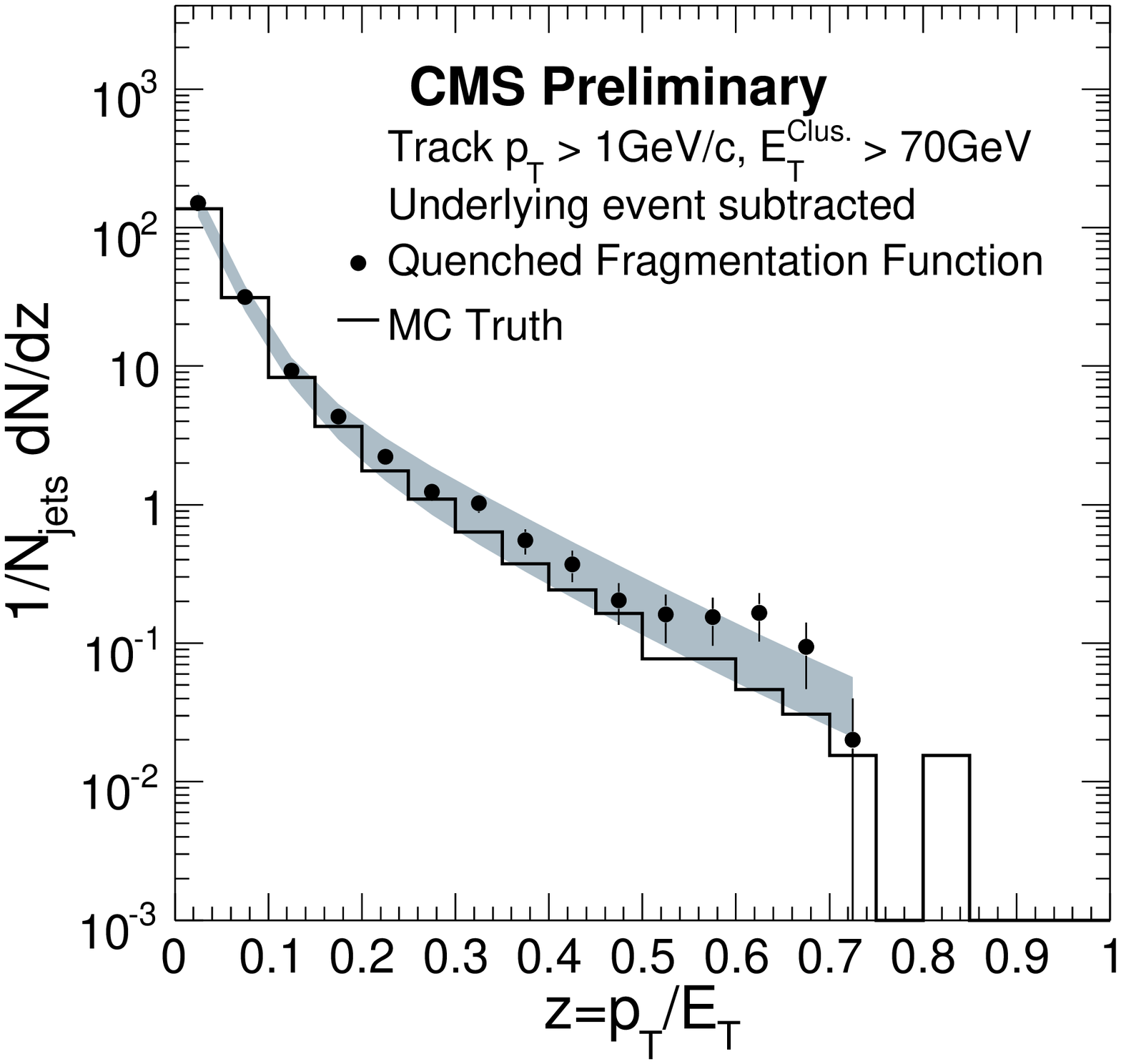}
\includegraphics[height=0.374\textwidth]{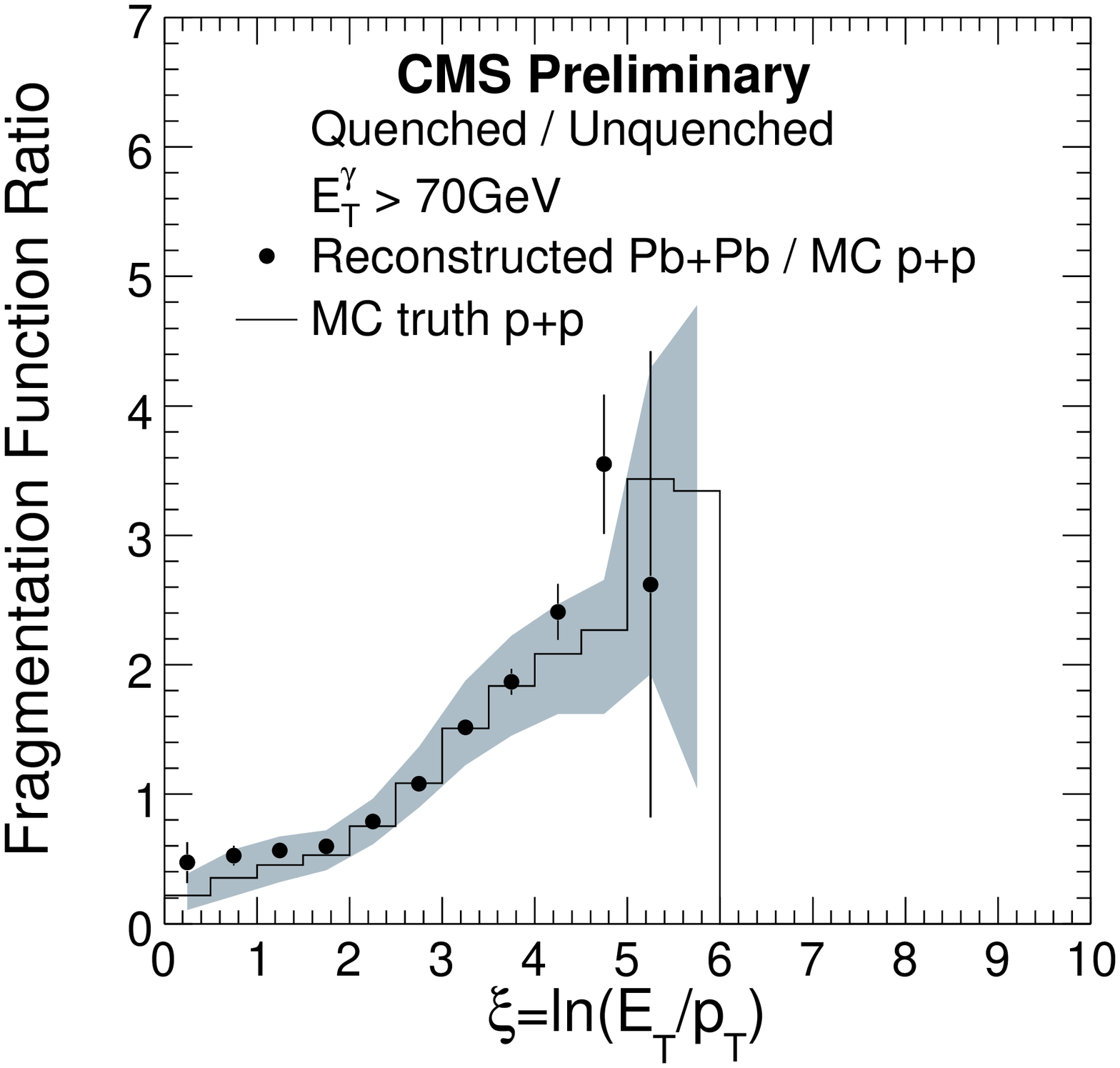}
\caption[]{
Left: reconstructed (symbols) and truth (histogram) fragmentation function of 
quenched jets with $E_T^\gamma>70$~GeV in Pb+Pb collisions at 
$\sqrt{s_{_{\rm NN}}}=5.5$~TeV. Right: reconstructed (symbols) and truth 
(histogram) ratio of quenched and non-quenched fragmentation functions.
The gray bands show the systematic errors on the measurement.}
\label{fragfunc}
\end{figure}

An example for the reconstructed fragmentation function of quenched jets after 
subtraction of the underlying event is shown on the left panel of
Fig.~\ref{fragfunc}, for the sample where the minimum $E_T$ of the
photon candidate ECAL cluster was 70 GeV. Also shown in Fig.~\ref{fragfunc}
the ratio of quenched and non-quenched fragmentation function after reconstruction 
(symbols) and the simulated truth (histogram). The estimated systematic errors (gray 
band) are small enough to allow the quantitative measurement of the medium 
modification of the jet fragmentation function~\cite{gammajet_pas}.

%%\section{Summary}

To summarize, some of the capabilities of CMS for high-$p_T$ studies were presented. 
In particular, 
an overview of the tracking, jet, photon and dimuon reconstruction and triggering
was given. Using $\gamma$-jet events, the expected strong modification of jet 
fragmentation function can be measured. With the above capabilities, a whole list of 
new observables will become available to CMS at the LHC, like dijet correlations, 
$Z^0$-jet events, b-tagged jets, forward jets, among others.

%% end of main text

%\section*{Acknowledgments} % please insert, comment out or delete if not needed
%This is where one places acknowledgments for funding bodies etc., if needed.
%For the large collaborations, this is listed once and for all, together with 
%the author lists etc. in the proceedings back-material.

%%------------------------------------------------------------------
 % do not change 

\end{document}